\documentstyle[psfig,11pt,aaspp4]{article}

\def\etal{et al.\,}
\def\Bruntfreq{Brunt-V{\"a}is{\"a}l{\"a}\,\,}
\def\refnew#1{(\ref{#1})}
\def\be{\begin{equation}}
\def\ee{\end{equation}}

\def\vcv{v_{\rm cv}}
\def\tcv{t_{\rm cv}}

\def\erg{\, \rm erg}
\def\K{\, \rm K}
\def\s{\, \rm s}

\def\cm{\, \rm cm}
\def\gm{\, \rm gm}
\def\dyne{\, \rm dyne}

\def\ph{\rm{ph}}
\def\th{\rm {th}}
\newcommand{\bxi}{\mbox{\boldmath $\xi$}}
\newcommand{\bnabla}{\mbox{\boldmath $\nabla$}}

\begin{document} 

\title{\mbox{GRAVITY-MODES IN ZZ CETI STARS} \\ \mbox I. 
QUASIADIABATIC ANALYSIS OF OVERSTABILITY}
\author{Peter Goldreich\altaffilmark{1} and Yanqin Wu\altaffilmark{1,2}}

\altaffiltext{1}{Theoretical Astrophysics, California Institute of Technology 
	130-33, Pasadena, CA 91125, USA; pmg@gps.caltech.edu}
\altaffiltext{2}{Astronomy Unit, School of Mathematical Sciences, 
	 Queen Mary and Westfield College, Mile End Road, London E1 4NS, UK;
                 Y.Wu@qmw.ac.uk}

\begin{abstract}
We analyze the stability of g-modes in variable white dwarfs with
hydrogen envelopes. All the relevant physical processes take place in
the outer layer of hydrogen rich material which consists of a
radiative layer overlain by a convective envelope. The radiative layer
contributes to mode damping because its opacity decreases upon
compression and the amplitude of the Lagrangian pressure perturbation
increases outward. The convective envelope is the seat of mode
excitation because it acts as an insulating blanket with respect to
the perturbed flux that enters it from below.  A crucial point is that
the convective motions respond to the instantaneous pulsational state.
Driving exceeds damping by as much as a factor of two provided 
$\omega\tau_c\geq 1$, where $\omega$ is the radian frequency of the
mode and $\tau_c\approx 4\tau_{\rm th}$ with $\tau_{\rm th}$ being the
thermal time constant evaluated at the base of the convective
envelope. As a white dwarf cools, its convection zone deepens, and
modes of lower frequency become overstable.  However, the deeper
convection zone impedes the passage of flux perturbations from the
base of the convection zone to the photosphere. Thus the photometric
variation of a mode with constant velocity amplitude decreases. These
factors account for the observed trend that longer period modes are
found in cooler DAVs. Overstable modes have growth rates of order
$\gamma\sim 1/(n\tau_\omega)$, where $n$ is the mode's radial order
and $\tau_\omega$ is the thermal time-scale evaluated at the top of
the mode's cavity. The growth time, $\gamma^{-1}$, ranges from hours
for the longest period observed modes ($P\approx 20$ minutes) to
thousands of years for those of shortest period ($P\approx 2 $
minutes).  The linear growth time probably sets the time-scale for
variations of mode amplitude and phase.  This is consistent with
observations showing that longer period modes are more variable than
shorter period ones.  Our investigation confirms many results obtained
by Brickhill in his pioneering studies of ZZ Cetis. However, it
suffers from at least two serious shortcomings. It is based on the
quasiadiabatic approximation that strictly applies only in the limit
$\omega\tau_c\gg 1$, and it ignores damping associated with turbulent
viscosity in the convection zone. We will remove these shortcomings in
future papers.
\end{abstract}


\setcounter{equation}{0}

\section{Introduction\label{sec:adia-intro}}

ZZ Cetis, also called DAVs, are variable white dwarfs with hydrogen
atmospheres.  Their photometric variations are associated with
nonradial gravity-modes (g-modes); for the first conclusive proof, see
Robinson \etal (\cite{scaling-robinson82}). These stars have shallow
surface convection zones overlying stably stratified interiors. As the
result of gravitational settling, different elements are well
separated . With increasing depth, the composition changes from
hydrogen to helium, and then in most cases to a mixture of carbon and
oxygen. From center to surface the luminosity is carried first by
electron conduction, then by radiative diffusion, and finally by
convection.

Our aim is to describe the mechanism responsible for the overstability
of g-modes in ZZ Ceti stars. This topic has received attention in the
past. Initial calculations of overstable modes were presented in
Dziembowski \& Koester (\cite{adia-dziem81}), Dolez \& Vauclair
(\cite{adia-dolez81}), and Winget \etal (\cite{adia-winget82}). These
were based on the assumption that the convective flux does not respond
to pulsation; this is often referred to as the frozen convection
hypothesis.  Because hydrogen is partially ionized in the surface
layers of ZZ Ceti stars, these workers attributed mode excitation to
the $\kappa$-mechanism. In so doing, they ignored the fact that the
thermal time-scale in the layer of partial ionization is many orders
of magnitude smaller than the periods of the overstable modes. Pesnell
(\cite{adia-pesnell87}) pointed out that in calculations such as those
just referred to, mode excitation results from the outward decay of
the perturbed radiative flux at the bottom of the convective
envelope. He coined the term `convective blocking' for this excitation
mechanism.\footnote{This mechanism was described in a general way by
Cox \& Guili (\cite{adia-cox68}), and explained in more detail by
Goldreich \& Keeley (\cite{adia-goldreich77}).} Although convective
blocking is responsible for mode excitation in the above cited
references, it does not occur in the convective envelopes of ZZ Ceti
stars. This is because the dynamic time-scale for convective
readjustment (i.e., convective turn-over time) in these stars is much
shorter than the g-mode periods. Noting this, Brickhill
(\cite{adia-brick83}, \cite{adia-brick90}, \cite{adia-brick91a},
\cite{adia-brick91b}) assumed that convection responds instantaneously
to the pulsational state. He demonstrated that this leads to a new
type of mode excitation, which he referred to as convective
driving. Brickhill went on to presents the first physically consistent
calculations of mode overstability, mode visibility, and instability
strip width. Our investigation supports most of his
conclusions. Additional support for convective driving is provided by
Gautschy \etal (\cite{adia-gautschy96}) who found overstable modes in
calculations in which convection is modeled by hydrodynamic
simulation.

In this paper we elucidate the manner in which instantaneous
convective adjustment promotes mode overstability. We adopt the
quasiadiabatic approximation in the radiative interior. We also ignore
the effects of turbulent viscosity in the convection zone. These
simplifications enable us to keep our investigation analytical,
although we appeal to numerically computed stellar models and
eigenfunctions for guidance.  The DA white dwarf models we use are
those produced by Bradley (\cite{scaling-bradley96}) for
asteroseismology.  Fully nonadiabatic results, which require numerical
computations, will be reported in a subsequent paper. These modify the
details, but not the principal conclusions arrived at in the present
paper.

The plan of our paper is as follows. The linearized wave equation is
derived in \S \ref{sec:adia-prepare}. In \S \ref{sec:adia-perturb}, we
evaluate the perturbations associated with a g-mode in different parts
of the star. We devote \S \ref{sec:adia-driving} to the derivation of
a simple overstability criterion. Relevant time-scales and the
validity of the quasiadiabatic approximation are discussed in \S
\ref{sec:adia-discussion}. The appendix contains derivations of
convenient scaling relations for the dispersion relation, the WKB
eigenfunction, and the amplitude normalization.

\section{Preparations}
\label{sec:adia-prepare}

We adopt standard notation for the thermodynamic variables pressure,
$p$, density, $\rho$, and temperature, $T$. Our specific entropy, $s$,
is dimensionless; we measure it in units of $k_B/m_p$, where $k_B$ is
the Boltzmann constant and $m_p$ the proton mass. We denote the
gravitational acceleration by $g$, the opacity per unit mass by
$\kappa$, the pressure scale height by, $H_p
\equiv p/(g\rho)$, and the adiabatic sound speed by $c_s$ where 
$c_s^2 \equiv ({\partial p} /{\partial \rho})_s$.

A g-mode in a spherical star is characterized by three eigenvalues
$(n,\ell,m)$; $n$ is the number of radial nodes in the radial
component of the displacement vector, ${\ell}$ is the angular degree,
and $m$ the azimuthal separation parameter.\footnote{We ignore $m$ in
this paper since DAV's are slow rotators.} G-modes detected in DAV
stars have modest radial orders, $1\leq n\leq 25$, and low angular
degrees, $1\leq {\ell}\leq 2$.\footnote{The latter is probably an
observational selection effect.}.  Their periods fall in the range
between $100 \s$ and $1200 \s$. G-mode periods increase with $n$ and
decrease with ${\ell}$ (see \S
\ref{subsec:adia-scaling-disp}). These properties reflect the nature of the 
restoring force, namely gravity which opposes departure of surfaces
of constant density from those of constant gravitational potential.

Gravity waves propagate where $\omega$ is smaller than both the radian
Lamb (acoustic) frequency, $L_\ell$, and the radian \Bruntfreq
(buoyancy) frequency, $N$. These critical frequencies are defined by
$L_{\ell}^2={{\ell}({\ell}+1)}(c_s/r)^2$, and $N^2\equiv -
g(\partial\ln\rho/\partial s)_p \, ds/dr$.  Electron degeneracy
reduces the buoyancy frequency in the deep interior of a white dwarf
(cf. Fig. 1). Consequently, the uppermost radial node of even the
lowest order g-mode is confined to the outer few percent of the
stellar radius.  Since all of the relevant physics takes place in the
region above this node, it is convenient to adopt a plane-parallel
approximation with $g$ taken to be constant in both space and time.
In place of the radial coordinate $r$, we use the depth $z$ below the
photosphere. We neglect gravitational perturbations; due to the small
fraction of the stellar mass that a mode samples, its gravitational
perturbation is insignificant.  A few order of magnitude relations to
keep in mind are: $p\sim g\rho z$, $H_p\sim z$, $d\ln\rho/dz\sim 1/z$,
and $c_s^2\sim gz$. In the convection zone, $N^2\sim -(\vcv/c_s)^2 g/z
\sim - (\vcv/z)^2$, where $\vcv$ is the convective velocity. In the
upper radiative interior, $N^2\sim g/z \sim (c_s/z)^2$.

The linearized equations of mass and momentum conservation, augmented
by the linearized, adiabatic equation of state, read\footnote{Readers
are referred to Unno \etal (\cite{scaling-unno89}) for detailed
derivations of these equations.}
\begin{eqnarray}
{\delta\rho\over\rho} & = & -ik_h\xi_h-{d\xi_z\over dz},
\label{eq:adia-cont} \\ 
\omega^2\xi_h & = & ik_h\left[{p\over\rho} \left({\delta p\over
p}\right)-g\xi_z\right], \label{eq:adia-momh} \\
\omega^2\xi_z & = & {p\over\rho}{d\over dz}\left({\delta p\over p}\right)
+g\left({\delta p\over p}-{\delta\rho\over\rho}-{d\xi_z\over dz}\right),
\label{eq:adia-momz} \\
\delta p & = & c_s^2\delta\rho. \label{eq:adia-state}
\end{eqnarray}
In the above equations, $\xi_h$ and $\xi_z$ represent horizontal and
vertical components of the displacement vector, $\delta$ denotes a
Lagrangian perturbation, and $\exp{i(k_h x -\omega t)}$ describes the
horizontal space and time dependence of the perturbations. The
horizontal component of the propagation vector, $k_h$, is related to
the angular degree, ${\ell}$, by
\begin{equation}
k_h^2={{\ell}({\ell}+1)\over R^2}, \label{eq:adia-kh}
\end{equation}
where $R$ is the stellar radius. After some manipulation, equations
\refnew{eq:adia-momh} and
\refnew{eq:adia-momz} yield
\begin{equation}
\xi_h={ig^2k_h\over (gk_h)^2-\omega^4}\left[{p\over g\rho}{d\over
dz}\left({\delta p\over p}\right)+\left(1-{\omega^2p\over
g^2\rho}\right)
\left({\delta p\over p}\right)\right], \label{eq:adia-xih}
\end{equation}
and 
\begin{equation}
\xi_z={-g\omega^2\over (gk_h)^2-\omega^4}\left[{p\over g\rho}{d\over
dz}\left({\delta p\over p}\right)+\left(1-{k_h^2p\over\omega^2\rho}
\right)\left({\delta p\over p}\right)\right]. \label{eq:adia-xiz}
\end{equation}
Notice that these two equations are independent of the assumption of
adiabaticity.

The linear, adiabatic wave equation for the fractional, Lagrangian
pressure perturbation, $\delta p/p$, follows from combining equations
\refnew{eq:adia-cont}, \refnew{eq:adia-xih}, and \refnew{eq:adia-xiz},
\begin{equation}
{d^2\over dz^2}\left({\delta p\over p}\right)+{d\over
dz}\left(\ln{p^2\over\rho}\right){d\over dz}\left({\delta p\over
p}\right)+
\left[k_h^2\left({N^2\over\omega^2}-1\right)+\left({\omega\over
c_s}\right)^2 \right]\left({\delta p\over
p}\right)=0. \label{eq:adia-wave}
\end{equation}
The advantages of choosing $\delta p/p$ as the dependent variable will
become apparent as we proceed.

The appendix is devoted to deducing the properties of g-modes from the
wave equation \refnew{eq:adia-wave}. Here we summarize results that
are needed in the main body of the paper.  A g-mode's cavity coincides
with the radial interval within which $\omega$ is smaller than both
$N$ and $L_{\ell}$. Inside this cavity the mode's velocity field is
relatively incompressible.  The lower boundary of each g-mode's cavity
is set by the condition $\omega=N$.  For modes of sufficiently high
frequency, the relation $\omega=L_\ell$ is satisfied in the radiative
interior, and it determines the location of the upper
boundary. Otherwise, the upper boundary falls in the transition layer
between the radiative interior and the convective envelope where
$\omega=N$. In this paper, we restrict consideration to modes which do
not propagate immediately below the convection zone since these
include the overstable modes. To order of magnitude, the condition
$\omega \sim L_{\ell}$ is satisfied at $z_\omega \sim
{\omega^2}/{(gk_h^2)}$. If we denote the depth of the first radial
node of ${\delta p/p}$ as $z_1$, and the depth at the bottom of the
surface convection zone as $z_b$, then $z_1 \sim z_\omega$ provided
$z_\omega > z_b$.

The fractional Lagrangian pressure perturbation, $\delta p/p$, is
almost constant in the convection zone ($z\leq z_b$), and declines
smoothly to zero in the upper evanescent region ($ z_b \leq z \leq
z_1$). Its envelope scales with depth in the cavity so as to maintain
a constant vertical energy flux.  An equal amount of energy is stored
between each pair of consecutive radial nodes. For modes having
$z_\omega>z_b$, the photospheric value of the normalized eigenfunction
is given by $(\delta p/p)_{\ph}\sim 1/(n\tau_\omega L)^{1/2}$.

\begin{figure}
\centerline{\psfig{figure=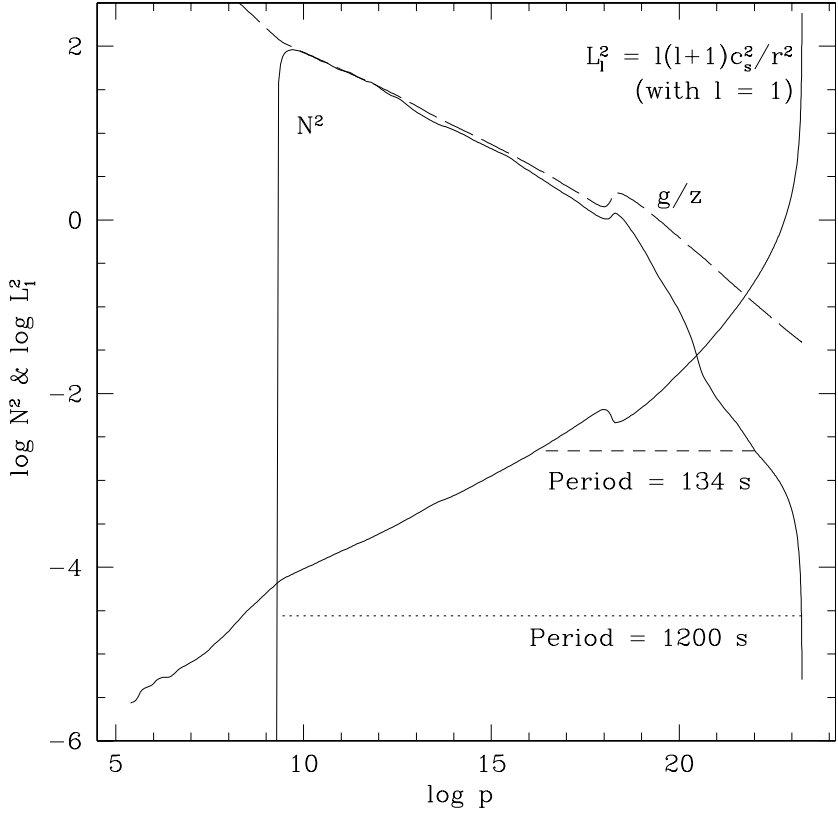,width=0.75\hsize}}
\caption[]{The squares of the radian \Bruntfreq frequency, $N^2$, and
Lamb frequency, $L_{\ell}^2$, in $\s^{-2}$ as functions of pressure in
$\dyne\cm^{-2}$ for Bradley's white dwarf model with $T_{\rm
eff}=12,000\K$.  $N^2$ is negative in the convection zone $(\log p\leq
9.3)$, and drops with increasing depth to near zero in the degenerate
interior $(\log p\geq 23)$.  The little bump at $\log p
\sim 19$ is due to the sharp compositional transition from hydrogen to
helium. The long dashed line is our analytic approximation, $N^2 \sim
(g/z)$. Notice that it worsens as degeneracy becomes important.
G-modes with frequency $\omega$ propagate in regions where both
$\omega < N$ and $\omega < L_{\ell}$. These regions are depicted by
horizontal dashed lines for two ${\ell} = 1$ modes having periods of
$134 \s$ and $1200 \s$ respectively. The former has its upper cavity
terminated where $\omega = L_{\ell}$, while the latter propagates
until immediately below the convection zone where $\omega =N$.
\label{fig:scaling-bruntplot}}
\end{figure}

\section{Perturbations Associated with G-Mode Pulsations
\label{sec:adia-perturb}}

Here we derive expressions relating various perturbation quantities
associated with g-modes.\footnote{When reading this section, it is
important to bear in mind that all perturbation quantities have an
implicit time and horizontal space dependence of the form $\exp{i(k_h
x-\omega t)}$.} These are used in \S
\ref{sec:adia-driving} to evaluate
the driving and damping in various parts of the star. Since
theoretical details are of limited interest, we summarize selected
results at the end of this section.

We start with the radiative interior and proceed outward through the
convection zone to the photosphere. A new symbol, $\Delta$, is
introduced to denote variations associated with a g-mode at a
particular level, such as the photosphere, or within a particular
layer, such as the convection zone.  These variations are not to be
confused with either Lagrangian variations which we denote by
$\delta$, or Eulerian variations.

\begin{figure}
\centerline{\psfig{figure=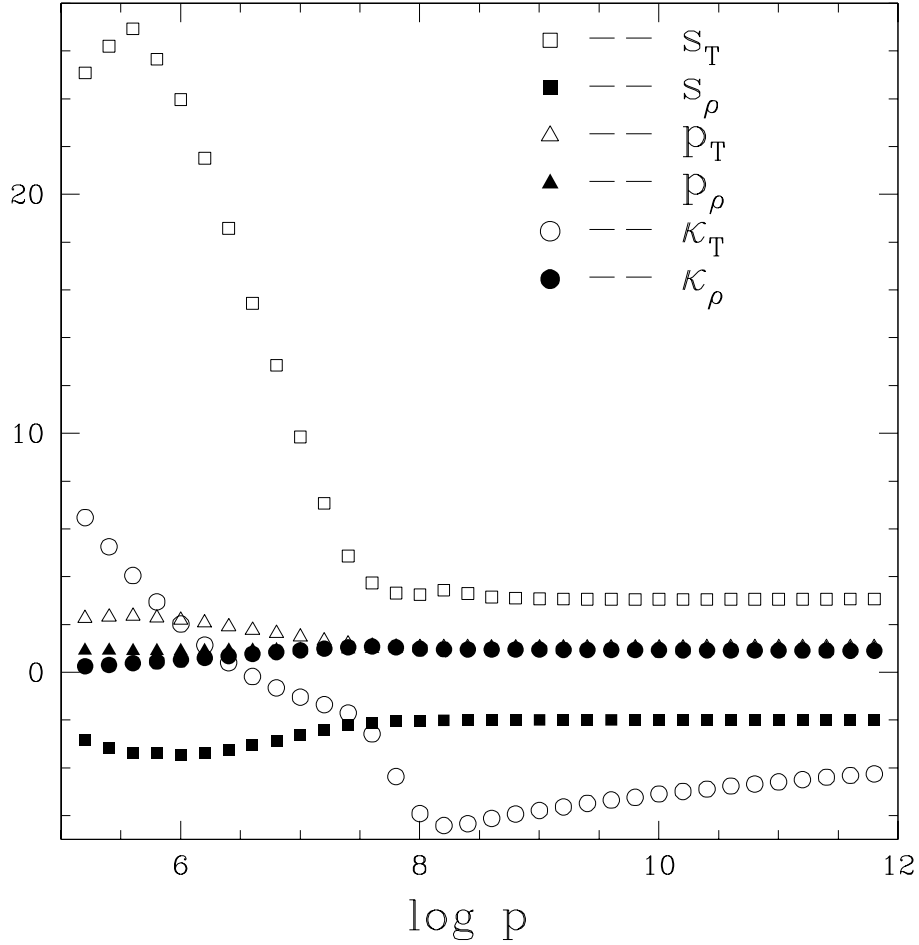,width=0.85\hsize}}
\caption[]{Values of various dimensionless thermodynamic quantities 
as a function of pressure in $\dyne\cm^{-2}$ for Bradley's white dwarf
model with $T_{\rm eff}=12,300\K$. The convection zone extends from
the surface down to $\log p \sim 8.5$. Both $s_T$ and $\kappa_T$
change significantly with depth in the upper layers where hydrogen is
partially ionized. The fully ionized region has mean molecular weight
$\mu = 1/2$. Other thermodynamic quantities can be obtained from those
plotted by using the relations: $\rho_s = p_T/(p_T s_\rho - p_\rho
s_T)$, $c_p = (s_T p_\rho - s_\rho p_T)/p_\rho$, and $\Gamma_1 = c_s^2
\rho/p = (p_\rho s_T - p_T s_\rho)/s_T$.
\label{fig:adia-scaling-sTandsoon} }
\end{figure}

This section is replete with thermodynamic derivatives. In most
instances we take $\rho$ and $T$ as our independent variables. Unless
specified otherwise, it is implicitly assumed that partial derivatives
with respect to one are taken with the other held constant. We adopt
the shorthand notation: $p_\rho\equiv \partial\ln p/\partial\ln \rho$,
$p_T\equiv \partial \ln p/\partial \ln T$, $s_\rho\equiv\partial
s/\partial\ln\rho $, $s_T\equiv\partial s/\partial\ln T$,
$\kappa_\rho\equiv
\partial\ln\kappa/\partial\ln \rho$, $\kappa_T\equiv \partial \ln
\kappa/\partial \ln T$. However, we use the symbols
$\rho_s\equiv (\partial\ln\rho/\partial s)_p$ and $c_p\equiv
T(\partial s/\partial T)_p$.  These thermodynamic quantities evaluated
for a model DA white dwarf are displayed in Fig.
\ref{fig:adia-scaling-sTandsoon}.

\subsection{Radiative Interior\label{subsec:adia-radia}}

The radiative flux in an optically thick region obeys the diffusion
equation,
\begin{equation}
F={4\sigma\over 3\kappa\rho}{dT^4\over dz}, \label{eq:adia-Frad}
\end{equation}
where $\sigma$ is the Stefan-Boltzmann constant. The Lagrangian
perturbation of the flux takes the form
\begin{equation}
{\delta F\over F}=-(1+\kappa_\rho){\delta\rho\over\rho}
+(4-\kappa_T){\delta T\over T} - {d\xi_z\over dz} 
+\left({d\ln T\over dz}\right)^{-1}{d\over dz}
\left({\delta T\over T}\right). \label{eq:adia-eqdelF}
\end{equation}

Next we express $\delta F/F$ in terms of $\delta p/p$ within the
quasiadiabatic approximation. To accomplish this, we decompose $\delta
\rho/\rho$ and $\delta T/T$ into adiabatic and nonadiabatic components
by means of the thermodynamic identities
\begin{equation}
{\delta \rho\over \rho}={p\over c_s^2\rho}{\delta p\over p} + \rho_s\delta s 
\approx {3\over 5}{\delta p\over p}-{\delta s\over 5},
\label{eq:adia-eqdelrho}
\end{equation}
and 
\begin{equation}
{\delta T\over T}=-{s_\rho\over s_T}{p\over c_s^2\rho}{\delta p\over 
p} + {\delta s\over c_p}\approx {2\over 5}{\delta p\over p}+{\delta s\over 5},
\label{eq:adia-eqdelT}
\end{equation}
where the coefficients are evaluated as appropriate for fully ionized
hydrogen plasma (cf. Fig. \ref{fig:adia-scaling-sTandsoon}).
Substitution of the adiabatic parts of these expressions into equation
\refnew{eq:adia-eqdelF} is carried out separately for the upper evanescent layer 
and the g-mode cavity.

\subsubsection{Upper Evanescent Layer}

Within the region $z_b\leq z \leq z_\omega$, ${\delta p/p}\approx -i
k_h \xi_h$, and it varies on spatial scale $z_\omega$ (see the
appendix). Thus equation (\ref{eq:adia-cont}) leads to
\begin{equation}
{d\xi_z\over dz}\approx {\delta p\over p}-{\delta \rho\over \rho}.
\label{eq:adia-eqdxizdz} 
\end{equation}
Moreover, the term 
\begin{equation}
\left({d\ln T\over dz}\right)^{-1}{d\over dz}\left({\delta T\over
T}\right) \approx -\left({d\ln T\over dz}\right)^{-1}{d\over
dz}\left({s_\rho\over s_T}{p\over c_s^2\rho}{\delta p\over
p}\right), \label{eq:adia-graddelT}
\end{equation}
is of order $(z/z_\omega)(\delta p/p)$ and can be
discarded.\footnote{As shown in Fig. \ref{fig:adia-scaling-sTandsoon},
the factors $s_\rho\approx -2$, $s_T\approx 3$, and
$p/c_s^2\rho\approx 3/5$ are each nearly constant in the upper part of
the radiative interior.} Making these approximations, we arrive at
\begin{equation}
{\delta F\over F} \approx A {{\delta p}\over p},
\label{eq:adia-eqdelFup} 
\end{equation}
where 
\begin{equation}
A={\left(3-3\kappa_\rho-2\kappa_T\right)\over 5}.
\label{eq:adia-A}\end{equation}

\subsubsection{G-Mode Cavity}

For $z\geq z_\omega$, perturbations vary on the spatial scale of
$1/k_z$, where $k_z\approx (N/\omega)k_h>1/z$ is the vertical
component of the wave vector as given by equation \refnew{eq:adia-DR}.
Here, the dominant contributions to $\delta F/F$ come from the last
two terms in equation (\ref{eq:adia-eqdelF}). The first transforms to
\begin{equation}
{d\xi_z\over dz}\approx -ik_z{\delta p\over g\rho}\sim -ik_z z{\delta p\over p}, 
\label{eq:adia-eqdxizdzl} 
\end{equation}
as most easily seen from equation \refnew{eq:adia-momz}. It then follows
that
\begin{equation}
{\delta F\over F}\approx {ik_z p\over g\rho}{d\ln p\over d\ln
T}\left({\partial \ln T\over\partial\ln p}\biggr\vert_s-{d\ln T\over
d\ln p}\right){\delta p\over p}\approx {ik_z N^2\over g}\left({p\over
g\rho}\right)^2 {d\ln p\over d\ln T}{\delta p\over p},
\label{eq:adia-eqdelFlp} 
\end{equation}
where we have set $p_\rho\approx 1$ and $p_T\approx 1$ as is appropriate for
a fully ionized plasma (cf. Fig. \ref{fig:adia-scaling-sTandsoon}).

\subsection{Convective Envelope\label{subsec:adia-convec}} 

The net absorption of heat by the convective envelope is given by
\begin{equation}
\Delta Q=\int_{\rm cvz}\, dz\, \rho{k_B\over m_p}T\, \delta s\approx
\Delta s_b\int_{\rm cvz}\, dz\, \rho{k_B\over m_p}T. \label{eq:adia-eqDelQ}
\end{equation}
Here $\Delta s_b$ is the variation of the specific entropy evaluated
at the bottom of the convection zone; it is neither a Lagrangian nor
an Eulerian variation.  The chain of argument leading to the second
expression for $\Delta Q$ requires some discussion. At equilibrium (no
pulsation), convection is so efficient that $ds/dz\ll s/z$ except in a
thin superadiabatic layer just below the photosphere; the unperturbed
convection zone is nearly isentropic. Moreover, the rapid response of
the convective motions ensures that $d\delta s/dz\ll
\delta s/z$; the vertical component of the perturbed entropy gradient
is small during pulsation. Because both $ds/dz\approx 0$ and $d\delta
s/dz\approx 0$, $\Delta s_b\approx \delta s$ .

To relate $\delta s$ with the flux perturbation, we treat convection
by means of a crude mixing-length model.  At equilibrium, the
convective flux is set equal to\footnote{Where the convection is
efficient, $F$ is comparable to the total flux.}
\begin{equation}
F\sim \alpha H_p\vcv {\rho k_B T\over m_p}{ds\over dz}, \label{eq:adia-Fcv}
\end{equation}
where $\alpha$ is the ratio of mixing length to the pressure scale height. The 
convective velocity, $\vcv$, satisfies\footnote{Note that $\rho_s
\equiv p_T/ (p_Ts_\rho-p_\rho s_T) < 0$.}
\begin{equation}
\vcv^2\sim -(\alpha H_p)^2 g\rho_s {ds\over
dz}=-\left(\alpha  H_p N\right)^2. \label{eq:adia-vcv}
\end{equation}
Eliminating $ds/dz$ and solving for $\vcv$, we find 
\begin{equation}
F\sim \rho\vcv^3.
\label{eq:adia-Fvcv}\end{equation}
Reversing the procedure and solving for $ds/dz$ yields
\begin{equation}
s_b-s_{\rm ph}\equiv \int_{\rm cvz}\, dz{ds\over dz} \approx H_p {ds\over dz}
\approx f\left({-1\over\rho_s\rho p}\right)^{1/3}
\left({m_pF\over k_B T}\right)^{2/3}, \label{eq:adia-eqsmsb} 
\end{equation}
where $s_{\rm ph}$ is the entropy at the photosphere.  The
dimensionless factor $f\sim \alpha^{-4/3}$ is of order unity.  The
right hand side of equation \refnew{eq:adia-eqsmsb} is evaluated at
the photosphere, since the entropy jump is concentrated immediately
below it.

We assume that the mixing length model applies during pulsation, as is
consistent with the short convective turn-over time.  Thus we write
the variation of $s_b-s_{\rm ph}$ associated with a g-mode as
\begin{equation}
{\Delta(s_b-s_{\rm ph})\over (s_b-s_{\rm ph})} = C {\Delta F_{\rm ph}\over F},
\label{eq:adia-eqDelsmsb} \end{equation}
where
\begin{eqnarray}
C & = & {1\over 12}
\biggr\{
6+{{p_T(1-\kappa_\rho) +\kappa_T(1+p_\rho)}\over {p_\rho+\kappa_\rho}}
+ {1\over {p_T(p_\rho s_T-p_Ts_\rho)}} 
\nonumber \\ 
& & \biggr [
p_T(p_{\rho T}s_T+p_\rho s_{\rm TT}-p_Ts_{\rho T})-p_T^2s_{\rho T}
-p_{\rm TT}p_\rho s_T 
\nonumber \\ 
& & +\left({p_T+\kappa_T}\over
{p_\rho+\kappa_\rho}\right)
\left(p_T^2s_{\rho\rho}-p_T(p_{\rho\rho}s_T+p_\rho s_{\rho T})+ p_\rho
p_{\rho T}s_T \right) \biggr ] \biggr\}. 
\label{eq:adia-C} \end{eqnarray}
The dimensionless number $C$ is to be evaluated at the photosphere.  In 
practice, only a few terms in the complicated expression for $C$ make 
significant contributions. In arriving at equations \refnew{eq:adia-eqDelsmsb}
and \refnew{eq:adia-C}, we implicitly assume that convection carries the entire 
stellar flux, and that $f$ does not vary during the pulsational cycle. 

\subsection{Photosphere\label{subsec:adia-photo}}

The photospheric temperature and pressure are determined by
\begin{equation}
T\approx\left({F\over \sigma}\right)^{1/4}, \label{eq:adia-Tphot}
\end{equation}
and
\begin{equation}
p\approx {2\over 3}{g\over \kappa}. \label{eq:adia-pphot} 
\end{equation}
Provided most of the flux in the photosphere is carried by radiation,
the temperature gradient is set by the equation of radiative diffusion
(eq. [\ref{eq:adia-Frad}]).\footnote{Applying the diffusion equation
at the photosphere is a crude approximation.}  Combining equations
\refnew{eq:adia-Frad}, \refnew{eq:adia-Tphot}, and
\refnew{eq:adia-pphot} yields
\begin{equation}
{d\ln T\over d\ln p}\approx {1\over 8}. 
\label{eq:adia-Tgrad} \end{equation}
On the other hand, the adiabatic gradient, 
\begin{equation}
{\partial \ln T\over\partial \ln p}\biggr\vert_s 
= {s_\rho\over p_T s_\rho -  p_\rho s_T}\approx {1\over 10},
\label{eq:adia-Tgradad} \end{equation}
using the numbers in Fig.
\ref{fig:adia-scaling-sTandsoon}. Comparing the photospheric and
adiabatic temperature gradients, we find that the convection zone
extends up to the photosphere for DA white dwarfs which lie inside the
instability strip. This is confirmed by comparison with numerical
models (e.g., Bradley
\cite{scaling-bradley96}, Wu \cite{scaling-phdthesis}). However, because the
photospheric density is low, convection is inefficient in this region, and 
radiation carries the bulk of the flux.

Next we relate changes in the thermodynamic variables at the
photosphere to changes in the emergent flux. Thus
\begin{equation}
{\Delta T\over T}\approx {1\over 4}{\Delta F_{\rm ph}\over F} 
\label{eq:adia-eqDetT}
\end{equation}
is an immediate consequence of equation (\ref{eq:adia-Tphot}). 
Expressions for $\Delta\rho/\rho$ and $\Delta p/p$ follow in a
straightforward manner from the thermodynamic identity,
\begin{equation}
{\Delta p\over p}\approx p_\rho {\Delta\rho\over \rho}+p_T{\Delta T\over T}, 
\label{eq:adia-idenprhoT}
\end{equation}
and from the perturbed form of equation (\ref{eq:adia-pphot}),
\begin{equation}
{\Delta p\over p}=-\kappa_\rho{\Delta\rho\over \rho} -
\kappa_T{\Delta T\over T}. \label{eq:adia-perphoto} 
\end{equation}
Together, these yield 
\begin{equation}
{\Delta\rho\over \rho}\approx -{1\over 4}\left({p_T+\kappa_T\over
p_\rho+\kappa_\rho}\right){\Delta F_{\rm ph}\over F}, \label{eq:adia-eqDelrho}
\end{equation}
and 
\begin{equation}
{\Delta p\over p}\approx {1\over 4}\left({p_T\kappa_\rho - p_\rho\kappa_T\over 
p_\rho+\kappa_\rho}\right) {\Delta F_{\rm ph}\over F}. \label{eq:adia-eqDelp} 
\end{equation}
It is then a simple step to show that 
\begin{equation}
\Delta s_{\rm ph}\approx B {\Delta F_{\rm ph}\over F},
\label{eq:adia-eqspert}
\end{equation}
where
\begin{equation}
B={1\over 4}\left[s_T-\left({p_T+\kappa_T\over
p_\rho+\kappa_\rho}\right)s_\rho\right].
\label{eq:adia-B}\end{equation} 

\subsection{Putting It All Together\label{subsec:adia-together}}

We begin by collecting a few key equations obtained in previous
sections.  These are then combined to determine how the entropy and
flux variations in the convection zone and photosphere depend upon the
pressure perturbation associated with the g-mode.

\subsubsection{Key Equations}

The flux perturbation entering the bottom of the
convective envelope from the radiative interior satisfies (eq. 
[\ref{eq:adia-eqdelFup}]),
\begin{equation}
{\Delta F_b\over F}\approx A\left({\delta p\over p}\right)_b.
\label{eq:adia-eqA}
\end{equation}
The photospheric entropy varies with the flux perturbation emerging
from the convection zone as (eq. [\ref{eq:adia-eqspert}])
\begin{equation}
\Delta s_{\rm ph} \approx B{\Delta F_{\rm ph}\over F}.
\label{eq:adia-eqDelsph} 
\end{equation}
The variation in the superadiabatic entropy jump is related to
this photospheric flux perturbation by (eq. [\ref{eq:adia-eqDelsmsb}])
\begin{equation}
\Delta (s_b-s_{\rm ph})\approx C{\Delta F_{\rm ph}\over F}.
\label{eq:adia-entjump} 
\end{equation}
The dimensionless numbers $A$, $B$ and $C$ are all positive; $A$ is evaluated at 
the boundary between the radiative interior and the convective envelope, whereas 
$B$ and $C$ are computed at the photosphere.

The net heat variation, $\Delta Q$, and specific entropy variation,
$\Delta s_b$, of the convective envelope are connected by equation
(\ref{eq:adia-eqDelQ}),
\begin{equation}
\Delta Q \approx  F\tau_b\Delta s_b, \label{eq:adia-eqDelQp} 
\end{equation}  
where the thermal time constant 
\begin{equation}
\tau_b\equiv {1\over F}\int_{\rm cvz}\, dz\, {\rho k_B T\over m_p}\approx 
{p_bz_b\over 7F}.
\label{eq:adia-taub}\end{equation}
The set of key equation is completed by the relation
\begin{equation}
{d\Delta Q\over dt}=\Delta F_b-\Delta F_{\rm ph}.
\label{eq:adia-dQdt}\end{equation}
We ignore horizontal heat transport for good reasons: transport by
radiation is completely negligible because $k_h z_b\ll 1$; turbulent
diffusion acts to diminish $\Delta s_b$, but only at the tiny rate
$k_h^2 z\,\vcv\ll \omega$.

For compactness of notation, we define a new thermal time constant, $\tau_c$, by
\begin{equation}
\tau_c\equiv (B+C)\tau_b. \label{eq:adia-eqtauc}
\end{equation}
The physical relation between $\tau_c$ and $\tau_b$ is discussed in
\S \ref{sec:adia-discussion}. Here we merely note that $\tau_c$ is
generally an order of magnitude or more larger than $\tau_b$.

\subsubsection{Implications Of Key Equations}

Taken together, the five homogeneous equations (\ref{eq:adia-eqA}),
(\ref{eq:adia-eqDelsph}), (\ref{eq:adia-entjump}),
(\ref{eq:adia-eqDelQp}), and (\ref{eq:adia-dQdt}) enable us to express
the five quantities $\Delta s_b$, $\Delta s_{\rm ph}$, $\Delta F_b/F$,
$\Delta F_{\rm ph}/F$, and $\Delta Q$ in terms of $(\delta p/
p)_b$.\footnote{We include $\Delta F_b/F$ in this list for
completeness although it is already expressed in terms of $(\delta
p/p)_b$ by equation (\ref{eq:adia-eqA}).}  The principal results from
this section read:
\begin{equation}
\Delta s_b\approx {A(B+C)\over 1-i\omega\tau_c}\left({\delta p\over p}\right)_b,
\label{eq:adia-Dsb} 
\end{equation}
\begin{equation}
\Delta s_{\rm ph}\approx {AB\over 1-i\omega\tau_c}\left({\delta p\over 
p}\right)_b, 
\label{eq:adia-Dsph} 
\end{equation}
\begin{equation}
{\Delta F_b\over F}\approx A\left({\delta p\over p}\right)_b, 
\label{eq:adia-DFb} 
\end{equation}
\begin{equation}
{\Delta F_{\rm ph}\over F}\approx {A \over 1-i\omega\tau_c}\left({\delta p\over 
p}\right)_b, \label{eq:adia-DFph} 
\end{equation}
\begin{equation}
\Delta Q \approx {AF\tau_c\over 1-i\omega\tau_c}\left({\delta p\over 
p}\right)_b. 
\label{eq:adia-DQ}
\end{equation}

\section{Driving and Damping\label{sec:adia-driving}}

The produce of this section is the proof that a g-mode is linearly
overstable in a ZZ Ceti star provided the convective envelope is thick
enough, or more precisely, provided $\omega\tau_c\gtrsim
1$.\footnote{The qualification that $z_\omega\geq z_b$ is necessary
here.}

The time-averaged rate of change in a mode's energy,
\begin{equation}
\gamma\equiv {\omega\over 2\pi}\oint dt {1\over E}{dE\over dt},
\label{eq:adia-work}
\end{equation}
is obtained from the work integral. Useful forms for $\gamma$ 
are\footnote{These assume that the eigenfunction is 
normalized such that $E=1$.} 
\begin{equation}
\gamma={2\omega R^2}\oint dt \int^R_0 dz\,\rho\,
{k_B\over m_p}\delta T\, {d\delta s\over dt} = {\omega\over 2\pi}L\oint
dt \int^R_0 dz {\delta T\over T}{d\over dz}\left({\delta F\over
F}\right).\label{eq:adia-workp}
\end{equation}
Regions of driving and damping are associated with positive and
negative values of the above integrand. The evaluation of $\gamma$ for
modes with $z_\omega \geq z_b$ is simplified by the near constancy of
$\delta p/p$ for $z\ll z_\omega$.

\subsection{Radiative Damping\label{subsec:adia-rdamping}}

In the upper evanescent region of the radiative interior, the perturbed flux 
varies in phase with the pressure perturbation because adiabatic compression 
causes the opacity to decrease.\footnote{$A > 0$ in equation 
\refnew{eq:adia-DFb}}. Since $\delta p/p$ declines from close to its 
surface value at $z_b$ to zero near $z_\omega$, and then oscillates
with rapidly declining amplitude at greater depth, the upper
evanescent layer loses heat during compression and thus contributes
most to mode damping.

To evaluate the radiative damping rate, it proves convenient to use the second 
form for $\gamma$ given by equation (\ref{eq:adia-workp}). 
We consider the contributions to $\gamma_{\rm rad}$ from the upper
evanescent layer, $\gamma_{r_u}$, and from the propagating cavity,
$\gamma_{r_l}$, separately.

The contribution from $z_b\leq z\leq z_\omega$, obtained with the aid
of equation (\ref{eq:adia-eqdelT}) for $\delta T/T$ and equation
(\ref{eq:adia-DFb}) for $\delta F/F$, reads
\begin{equation}
\gamma_{r_u}\approx - \left({A\, L\over 10}\right)\left({\delta
p\over p}\right)_b^2+{L\over 10}\int_{z_b}^{z_\omega} dz\left({\delta
p\over p}\right)^2{dA\over dz}. \label{eq:adia-Wdotru}
\end{equation}
The first term on the right hand side generally dominates over the
second one as $\delta p/p$ declines with depth and $A$ does not vary
significantly in this region.

For $z \geq z_\omega$, we again use equation \refnew{eq:adia-eqdelT} for 
$\delta T/T$, but now substitute equation (\ref{eq:adia-eqdelFlp}) for
$\Delta F/F$ to arrive at
\begin{equation}
\gamma_{r_l}\approx {-9k_h^2 L\over 125g^3\omega^2}
\int_{z_\omega}^\infty dz N^4 c^4 {d\ln p\over d\ln T} 
\left({\delta p\over p}\right)^2. \label{eq:adia-Wdotrl}
\end{equation}
Appeal to equation (\ref{eq:adia-Fluxp}) giving the WKB envelope
relation for $\delta p/p$ establishes that the integrand peaks close
to $z_\omega$.  Thus,
\begin{equation}
\gamma_{r_l}\sim -L\left({c N\over g}\right)_\omega^4
\left({\delta p\over p}\right)_\omega^2\sim -L\left({\delta p\over
p}\right)_\omega^2, 
\label{eq:adia-Wdotrlp}
\end{equation}
where in this context $(\delta p/p)_\omega$ represents the magnitude of the WKB 
envelope evaluated at $z_\omega$. Since this magnitude is significantly smaller 
than $(\delta p/p)_b$, the contribution to $\gamma$ from the g-mode cavity 
is negligible provided $z_\omega\geq z_b$. 

To a fair approximation, $\gamma_{\rm rad}$ may be set equal to the first
term on the right-hand side of equation (\ref{eq:adia-Wdotru}). Thus
\begin{equation}
\gamma_{\rm rad}\approx - \left({A\, L\over 10}\right)
\left({\delta p\over p}\right)_b^2. \label{eq:adia-Wdotr}
\end{equation}

\subsection{Convective Driving\label{subsec:adia-cdriving}}

The perturbed flux which exits the radiative interior enters the bottom of the 
convection zone. There it is almost instantaneously distributed so as to 
maintain the vertical entropy gradient near zero. Because the convection zone 
gains heat during compression, it is the seat of mode driving.

To evaluate the rate of convective driving we substitute equations 
(\ref{eq:adia-eqdelT}) and (\ref{eq:adia-Dsb})
into the first form for $\gamma$ given by equation
(\ref{eq:adia-workp}). It is apparent that the net contribution comes
entirely from the adiabatic part of $\delta T/T$ (see
eq. [\ref{eq:adia-eqdelT}]). Since the integrand is strongly weighted
toward the bottom of the convection zone, we evaluate all quantities there and 
arrive at
\begin{equation}
\gamma_{\rm cvz}\approx {(\omega \tau_c)^2 \over 1+ (\omega \tau_c)^2}
\left({A\, L\over 5}\right)\left({\delta p\over p}\right)_b^2. 
\label{eq:adia-Wdotcvz}
\end{equation}
Since $A$ is positive, so is $\gamma_{\rm cvz}$. 

\subsection{Turbulent Damping\label{subsec:adia-vdamping}}

Damping due to turbulent viscosity acting on the velocity shear in the
convection zone as well as the convective overshoot region is
estimated as
\begin{equation}
\gamma_{\rm visc}\approx -4\pi R^2\omega^2\int_0^{z_b}\, dz\, \rho\, \nu
\left({d\xi_h\over dz}\right)^2. \label{eq:adia-Edot}
\end{equation}
Here, $\nu\sim \vcv H_p $ is the turbulent viscosity, and 
$-i\omega d\xi_h/dz$ is the dominant component of velocity gradient.

Adiabatic perturbations in an isentropic region are
irrotational. Consequently, the velocity shear in an isentropic
convection zone would be very small and lead to negligible turbulent
damping. However, the mean entropy gradient and the horizontal
gradient of the entropy perturbation represent significant departures
from isentropy. In a future paper we demonstrate that turbulent viscosity 
suppresses the production of velocity gradients in the convection zone to the
extent that its effect on damping is negligible in comparison to
radiative damping.  However, turbulent damping in the region of
convective overshoot at the top of the radiative interior stabilizes
long period modes near the red edge of the ZZ Ceti instability strip.

\subsection{Net Driving\label{subsec:adia-net}}

The net driving rate follows from combining equations \refnew{eq:adia-Wdotr} and
\refnew{eq:adia-Wdotcvz};
\begin{equation}
\gamma_{\rm net}\approx {(\omega \tau_c)^2-1\over (\omega \tau_c)^2+1}
\left({A\, L\over 10}\right)\left({\delta p\over p}\right)_b^2.
\end{equation}
Driving exceeds damping by a factor of two in the limit $\omega\tau_c\gg 1$.
Substitution of the normalization relation given by equation 
\refnew{eq:adia-scaling-normdrho} for $\delta p/p$ yields the more revealing 
form
\begin{equation}
\gamma_{\rm net}\sim {(\omega \tau_c)^2-1\over 
(\omega \tau_c)^2+1}\left({1\over n\tau_\omega}\right). 
\label{eq:adia-Wdotnetp} 
\end{equation}
Overstability occurs if $\omega \tau_c > 1$.\footnote{In the
quasiadiabatic limit for modes with $z_\omega > z_b$.} Following
Brickhill, we refer to this excitation mechanism as convective
driving.

\section{Discussion\label{sec:adia-discussion}}

\subsection{Time-Scales \label{sec:adia-relevant}}

Three time-scales are relevant for convective driving in DAVs.  The
first is the period of an overstable g-mode, $P = 2\pi/\omega$, which
is typically of order a few hundred seconds. The second is the
dynamical time constant, $\tcv \sim H_p/\vcv$, on which convective
motions respond to perturbations; $\tcv \leq 1 \s$ throughout the
convection zones of even the coolest ZZ Cetis. This is why the
convective motions adjust to the instantaneous pulsational state. The
third is the thermal time constant, $\tau_c$, during which the
convection zone can bottle up flux perturbations that enter it from
below. 

Given the central role of $\tau_c$, we elaborate on its relation both
to $\tcv$ and to the more conventional definition of thermal time
constant, $\tau_{\rm th}$, at depth $z$. The latter is the heat
capacity of the material above that depth divided by the
luminosity. In a plane parallel, fully ionized atmosphere this is
equivalent to
\begin{equation}
\tau_{\rm th}\equiv {1\over F} \int_0^z dz \, c_p\, {\rho k_B\over m_p T}
\approx {5pz\over 7F}.
\label{eq:adia-tauth}\end{equation}
Appeal to equation \refnew{eq:adia-Fvcv} establishes that inside the
convection zone
\begin{equation}
{\tcv\over \tau_{\rm th}}\sim \left({\vcv\over c_s}\right)^2\ll 1.
\end{equation}
Now $\tau_c\equiv (B+C)\tau_b$, where $\tau_b$ is defined by equation
\refnew{eq:adia-taub}.  To the extent that $c_p\approx 5$ is constant in 
the convection zone, $\tau_b\approx \tau_{\rm th}/5$, where the latter
is evaluated at $z_b$.

Next we address the relation between $\tau_c$ and $\tau_b$. Here we
are concerned with the relatively large value of $B+C$, typically
about 20 for DAVs.\footnote{In our models, $B$ and $C$ have comparable
value.} Recall from equations \refnew{eq:adia-Dsb} and
\refnew{eq:adia-DFph} that
\begin{equation}
{\delta F_{\ph} \over F}\approx {\delta s_b\over B+C}.
\label{eq:adia-dFds}\end{equation}
So the photosphere and superadiabatic layer add an insulating blanket
on top of the convection zone.  The large value of $B$ follows because
the photospheres of DAVs are composed of lightly ionized hydrogen. In
this state, the values of $\kappa_T$ and $s_T$ are both large and
positive; typical values in the middle of the instability strip are
$\kappa_T\approx 6$ and $s_T\approx 24$.  The large and positive
$\kappa_T$ arises because the population of hydrogen atoms in excited
states which the ambient radiation field can photoionize increases
exponentially with increasing $T$. The large and positive $s_T$ occurs
as the ionization fraction increases exponentially with increasing
$T$, and the much larger entropy contributed by a free as compared to
a bound electron.  The large and positive value of $C$ reflects the
increase in entropy gradient that accompanies an increase in
convective flux. It is obtained from mixing length theory with an
unperturbed mixing length.

\subsection{Validity of the Quasiadiabatic Approximation}

The validity of the quasiadiabatic approximation requires that the
nonadiabatic parts of the expressions for $\delta \rho/\rho$ and
$\delta T/T$, as given by equations \refnew{eq:adia-eqdelrho} and
\refnew{eq:adia-eqdelT}, be small in comparison to the adiabatic
parts. Thus the ratio
\begin{equation}
{\cal R_{\rm na}}\equiv {\delta s\over\delta p/p}
\label{eq:adia-rnonad}\end{equation}
is a quantitative measure of nonadiabaticity. We estimate ${\cal R}_{\rm na}$ 
for the radiative interior and the convection zone.

We calculate $\delta s$ in the radiative interior from
\begin{equation}
\delta s \approx {iF\over \omega}{m_p\over\rho k_B T}{d\over dz}\left({\delta 
F\over F}\right).
\label{eq:adia-delsrad}\end{equation}
In the upper evanescent layer, $z_b<z<z_\omega$, this leads to
\begin{equation}
|{\cal R_{\rm na}}|\sim {1\over \omega\tau_{\th}},
\label{eq:adia-ratioevu}\end{equation}
whereas in the propagating cavity, $z>z_\omega$, we find
\begin{equation}
|{\cal R_{\rm na}}|\sim {1\over \omega\tau_{\th}}\left({z\over z_\omega}\right).
\label{eq:adia-ratioevl}\end{equation}
Nonadiabatic effects in the radiative interior are maximal at $z=z_b$, where
\begin{equation}
|{\cal R_{\rm na}}|\sim {1\over \omega\tau_b},
\label{eq:adia-ratioevmax}\end{equation}
since $\tau_{\rm th}/z$ increases with depth.  

The measure of nonadiabaticity in the convection zone is given by
equation \refnew{eq:adia-Dsb}. Since $\omega\tau_b>1$ is required for
the validity of the quasiadiabatic approximation in the radiative
zone, we restrict consideration to the limiting case $\omega\tau_c\gg
1$. In this limit, equation
\refnew{eq:adia-Dsb} yields
\begin{equation}
|{\cal R_{\rm na}}|\sim {1\over \omega\tau_b},
\label{eq:adia-ratiocvz}\end{equation}
which is identical to the value arrived at for the radiative zone in equation 
\refnew{eq:adia-ratioevmax}.

The requirement $\omega\tau_b\gtrsim 1$ for the validity of the quasiadiabatic 
approximation severely limits the applicability of the current investigation.
The perturbed flux at the photosphere is related to that at the bottom of the 
convection zone by
\begin{equation}
{\Delta F_{\ph}\over F}\approx {1\over 1-i\omega\tau_c}{\Delta F_b\over F}.
\label{eq:adia-relDelF}\end{equation}
Since $\tau_c$ is at least an order of magnitude larger than $\tau_b$, modes 
with $\omega\tau_b\gtrsim 1$ are likely to exhibit small photometric variations. 
However, this may not render them undetectable because their horizontal velocity
perturbations pass undiminished through the convection zone.

\subsection{Brickhill's Papers\label{subsec:adia-brickhill}}

Our investigation is closely related to studies of ZZ Cetis by
Brickhill
(\cite{adia-brick83},\cite{adia-brick90},\cite{adia-brick91a}). Brickhill
recognized that the convective flux must respond to the instantaneous
pulsational state. To determine the manner in which the convection
zone changes during a pulsational cycle, he compared equilibrium
stellar models covering a narrow range of effective
temperature. Brickhill provided a physical description of convective
driving and obtained an overstability criterion equivalent to
ours.\footnote{Our time constant $\tau_c$ is equivalent to the
quantity $D$ which Brickhill defined in equation (9) of his 1983
paper.} Moreover, he recognized that the convection zone reduces the
perturbed flux and delays its phase.

Our excuses for revisiting this topic are that Brickhill's papers are not widely
appreciated, that our approach is different from his, and that our paper 
provides the foundation for future papers which will examine issues beyond those
he treated.

\begin{appendix}

\section{Scaling Relations for Gravity-Modes\label{sec:adia-loworder}}

The appendix contains derivations of a number of scaling relations appropriate 
to g-modes in ZZ Cetis.  These relations are applied in the main
text and in our subsequent papers on white dwarf
pulsations. We obtain approximate expressions for dispersion relations,
eigenfunctions, and normalization constants.

\subsection{Properties of Gravity Waves}

Substituting the WKB ansatz 
\begin{equation}
{\delta p\over p}={\cal A}\exp\left({i\int^z\, dz\, k_z}\right),
\label{eq:adia-WKB}
\end{equation}
into the wave equation \refnew{eq:adia-wave}, and retaining terms of
leading order in $k_zH_p\gg 1$, we obtain the dispersion relation
\begin{equation}
k_z^2\approx
{\left(N^2-\omega^2\right)\left(k_h^2c_s^2-\omega^2\right)\over
c_s^2\omega^2},
\label{eq:adia-gwavedr}\end{equation}
where $k_hc_s$ is the plane parallel equivalent of the Lamb frequency, $L_\ell$.

Gravity waves propagate in regions where $\omega$ is smaller than both
$N$ and $L_\ell$.  Far from turning points, their WKB dispersion
relation simplifies to
\begin{equation}
k_z\approx \pm{N\over \omega}k_h.
\label{eq:adia-DR}\end{equation}
The vertical component of their group velocity is given by
\begin{equation}
v_{gz}\equiv {\partial \omega\over \partial k_z}\approx -{\omega\over
k_z}\approx \pm {\omega^2\over Nk_h}. 
\label{eq:adia-vgz} \end{equation}

Terms of next highest order in $k_zH_p$ yield the WKB amplitude
relation for gravity waves;
\begin{equation}
{\cal A}^2\propto {\rho\over k_z p^2}\propto {\rho\over N p^2}.
\label{eq:adia-amp}
\end{equation}
The amplitude relation expresses the conservation of the vertical flux
of wave action. This may be confirmed directly by means of an
alternate derivation. A straightforward manipulation of the linear
perturbation equations yields the quadratic conservation law
\begin{equation}
{\partial\over\partial t}\left({\rho\over 
2}\biggr\vert{\partial\bxi\over\partial t}\biggr\vert^2
+{(p^\prime)^2\over 2\rho c_s^2}+ {\rho\over 2}N^2\xi_z^2\right) = 
-\bnabla\cdot\left(p^\prime{\partial\bxi\over\partial t}\right),
\label{eq:adia-conserv}\end{equation}
where $p^\prime\equiv \delta p-g\rho\xi_z$ is the Eulerian pressure
perturbation.  We identify ${\cal{\bf F}}\equiv
p^\prime(\partial\bxi\partial t)$ as the quadratic energy flux. The
time-averaged magnitude of the vertical component of ${\cal{\bf F}}$,
computed to lowest order in $(k_z z)^{-1}$ with the aid of equations
\refnew{eq:adia-xiz} and \refnew{eq:adia-DR}, 
is found to be
\begin{equation}
{\cal F}\approx {\omega^2\over gk_h}{N p^2\over g\rho}\left({\delta
p\over p}\right)^2,
\label{eq:adia-Fz}\end{equation}
which is in accord with the WKB amplitude relation. Yet another way to
look at the energy flux is as the energy density transported by the
group velocity.  The energy density may be approximated as
$\omega^2\rho\xi_h^2$, since kinetic energy contributes half the
time-averaged total energy density, and $|\xi_h|\gg
|\xi_z|$. Application of equations \refnew{eq:adia-xih},
\refnew{eq:adia-DR}, and \refnew{eq:adia-vgz} yields
\begin{equation}
{\cal F}\approx \omega^2\rho\xi_h^2 v_{gz}\approx {\omega^2\over
gk_h}{N p^2\over g\rho}\left({\delta p\over p}\right)^2,
\label{eq:adia-Fluxp}\end{equation}
which reproduces the result for ${\cal F}$ given by equation
\refnew{eq:adia-Fz}.

In a spherical star, the conserved quantity is the WKB luminosity,
${\cal L}\equiv 4\pi r^2{\cal F}$. 

\subsection{Properties of G-Modes}

\subsubsection{Inside the Propagating Cavity\label{subsec:adia-cavity}}

The Lamb frequency, $L_\ell$, decreases with increasing $r$, whereas
the
\Bruntfreq frequency, $N$, increases until it drops abruptly just below the 
convection zone and becomes imaginary within it
(cf. Fig. 1). Compositional transitions may be exceptions to these
trends. The WKB approximation is violated if these transitions are
sharp on the scale of the radial wavelength. When these are located in
regions of propagation, they are best viewed as separating linked
cavities.

We divide a star's g-modes into two classes, high frequency modes and
low frequency modes. The former have cavities bounded from above at
$z_\omega>z_b$ where $\omega = L_\ell \approx k_h c_s$.  The latter
propagate until just below the convection zone at $z \approx z_b$
where $\omega=N$. The floor of the propagating cavity is set by
$\omega=N$ for all modes. In this paper we are exclusively concerned
with high frequency modes. Taking $c_s\sim (gz)^{1/2}$, we find
\begin{equation}
z_\omega\equiv {{\omega^2}\over{gk_h^2}}. \label{eq:adia-zone}
\end{equation}
Moreover, $N\sim (g/z)^{1/2}\sim c_s/z$ implies $k_z z_\omega\sim 1$ at
$z=z_\omega$. Thus $z_\omega$ is to be identified with $z_1$, the uppermost node 
of $\delta p/p$. More generally, 
\begin{equation}
k_z\sim {1\over \left(z_\omega z\right)^{1/2}},
\label{eq:adia-kz}\end{equation}
for $z>z_\omega$.

Expressions for $\xi_h$ and $\xi_z$ are given by equations (\ref{eq:adia-xih}) 
and (\ref{eq:adia-xiz}). To leading order in $k_z H_p$, these reduce to
\begin{equation}
\xi_h\approx -{k_z\over k_h}{p\over g\rho}\left({\delta p\over p}\right),
\label{eq:adia-xihc}
\end{equation}
and
\begin{equation}
\xi_z\approx  {p\over g\rho}\left({\delta p\over p}\right). 
\label{eq:adia-xizc} 
\end{equation}
Where $N/\omega\gg 1$, we have $k_z/k_h\gg 1$, which implies
$|\xi_h/\xi_z|\gg 1$. In these regions g-modes are characterized by
nearly horizontal motions.  Equations (\ref{eq:adia-xihc}) and
(\ref{eq:adia-xizc}) imply that $\delta
\rho/\rho = - ik_h\xi_h-d\xi_z/dz$ vanishes to leading order in
$k_zH_p\gg 1$. Thus g-modes are relatively incompressible within their
cavities.  A more precise estimate, obtained from equation
\refnew{eq:adia-xihc}, is $\delta\rho/\rho\sim
ik_h\xi_h(z_\omega/z)^{1/2}$.

\begin{figure}
\centerline{\psfig{figure=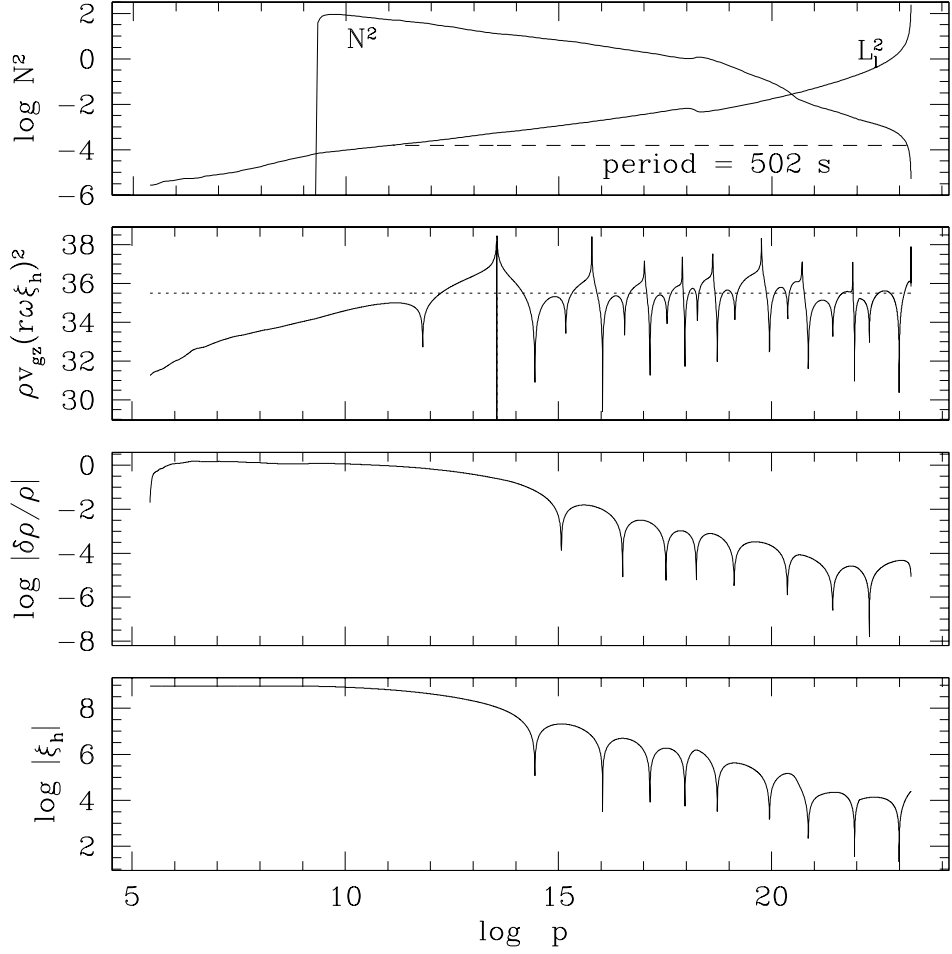,width=0.85\hsize}}
\caption[]{Structure of a mode with $n=9$, ${\ell}=1$, and a period 
of $502 \s$, as a function of pressure in $\dyne\cm^{-2}$. The top
panel, which is similar to Fig.
\ref{fig:scaling-bruntplot}, illustrates how the mode cavity, depicted
by the dashed line, is formed.  For this mode $z_\omega > z_b$.  The
WKB luminosity measured in $\erg /\s$ is plotted in the second panel.
It is constant inside the cavity and decays outside it. The
compositional transition between the hydrogen and helium layers has
minimal effect on this mode. The lower two panels display the depth
dependences for the dimensionless $\delta\rho/\rho$ and $\xi_h$
measured in $\cm$.  The numerical values in this figure come from
setting $\xi_h = R$ at the photosphere.}
\label{fig:adia-scaling-eigen}
\end{figure}

\subsubsection{Upper Evanescent Layer\label{subsec:adia-upper}}

For the purpose of this discussion, we pretend that both $\rho\to 0$
and $p\to 0$ as $z\to 0$. Then $z=0$ is a singular point of the wave
equation (\ref{eq:adia-wave}). Since the physical solution is regular
at $z=0$,
\begin{equation}
{d\over dz}\left[\ln\left({\delta p\over p}\right)\right]\sim 
-\left({N\over\omega}\right)^2 k_h^2 z. \label{eq:adia-reg} 
\end{equation} 
In the convection zone, where $N^2\approx -(\vcv/z)^2$, 
\begin{equation}
{d\over dz}\left[\ln\left({\delta p\over p}\right)\right]\sim
\left({\vcv\over c_s}\right)^2{1\over z_\omega}, \label{eq:adia-regcv}
\end{equation}
and at the top of the radiative interior, where $N^2\sim (c_s/z)^2$,
\begin{equation}
{d\over dz}\left[\ln\left({\delta p\over p}\right)\right]\sim 
-{1\over z_\omega}, \label{eq:adia-regr}
\end{equation}
We see that $\delta p/p$ is nearly constant for $z\ll z_\omega$ in the
upper evanescent layer.\footnote{That is why we chose to use it as the
dependent variable in the wave equation.} 

Taking into account the near constancy of $\delta p/p$, equations
(\ref{eq:adia-xih}) and (\ref{eq:adia-xiz}) imply that to leading
order in $z/z_\omega$,
\begin{equation}
\xi_h\approx {i\over k_h}\left({\delta p\over p}\right), 
\label{eq:adia-xihe} 
\end{equation}
and 
\begin{equation}
\xi_z\approx -{\omega^2\over gk_h^2}\left({\delta p\over p}\right). 
\label{eq:adia-xize} 
\end{equation}
Thus both components of the displacement vector are nearly constant for $z\ll 
z_\omega$.  The behavior of $\delta\rho/\rho$ is more subtle; in principle, it 
could vary on scale $z$ should $p/(c_s^2\rho)$ do so. However, in
practice, $p/(c_s^2\rho)$ exhibits only mild depth variations. Thus, 
equation (\ref{eq:adia-xihe}) yields
\begin{equation}
{\delta\rho\over\rho}\sim -ik_h\xi_h.
\label{eq:adia-incompp}
\end{equation} 
Equation (\ref{eq:adia-incompp}) shows that the relative incompressibility that
characterizes propagating g-modes does not extend to their evanescent
tails.

\subsection{Global Dispersion Relation \label{subsec:adia-scaling-disp}}

\begin{figure}
\centerline{\psfig{figure=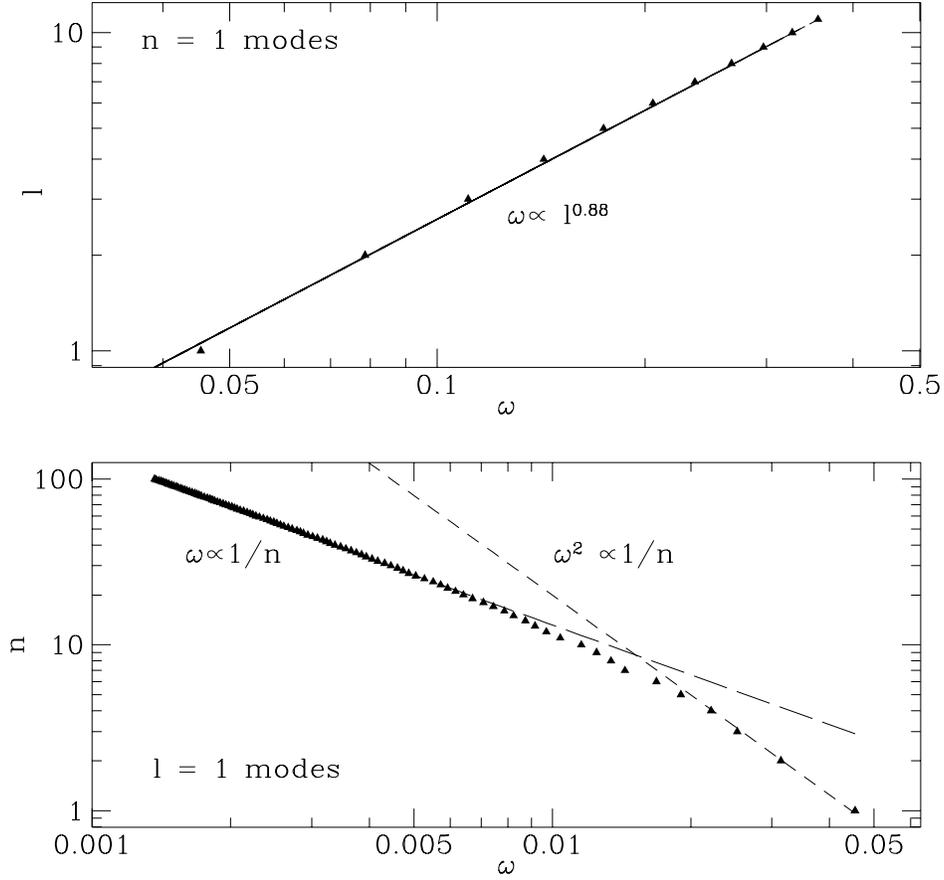,width=0.75\hsize}}
\caption[]{Radian frequencies of gravity-modes in $\s^{-1}$ as 
functions of radial order, $n$, and spherical degree, ${\ell}$.  These
eigenvalues are computed using Bradley's white dwarf model with
$T_{\rm eff} = 12,000
\K$. The upper panel is for $n = 1$ modes with various angular degrees,
while the lower panel is for ${\ell}= 1$ modes of different radial
orders. The global dispersion relations given by equations
\refnew{eq:adia-scaling-nl} and
\refnew{eq:adia-scaling-newnl} fit well for modes of low and high order 
respectively. We adopt these empirical laws in our analytical studies.
\label{fig:adia-scaling-disp} }
\end{figure}

The global dispersion relation is obtained from
\begin{equation}
n\pi\approx \int_{z_1}^{z_l} dz\, k_z,
\label{eq:adia-globalDR}
\end{equation}
where $z_l$ is the lower boundary of the mode cavity. Carrying out the
integration using $N^2\sim g/z$, $z_1 \sim z_\omega$, and equation
(\ref{eq:adia-DR}), we obtain $\omega^2\sim gk_h/n$.  This relation is
a good approximation to the g-mode dispersion relation for a
polytropic atmosphere. The proportionality
\be
\omega^2 \propto {{k_h}\over{n}}
\label{eq:adia-scaling-nl}
\ee
provides a satisfactory fit to the high frequency modes in DAV white
dwarfs.  However, low frequency modes penetrate deeply into the
interior where the approximation $N \sim (g/z)^{1/2}$ fails because of
electron degeneracy (cf. Fig. \ref{fig:scaling-bruntplot}).  As a
result of the steep drop in $N$, $z_l$ is nearly independent of
$\omega$ for $\omega \leq 10^{-2}\s^{-1}$. This steepens the
dependence of $\omega$ on $n$ such that
\be
\omega \propto {{k_h}\over{n}}.
\label{eq:adia-scaling-newnl}
\ee
Numerical results for the dispersion relations are shown in
Fig. \ref{fig:adia-scaling-disp}.

\subsection{Normalization of Eigenfunctions\label{subsec:adia-normal}}

We conform to standard practice and set
\begin{equation}
 {\omega^2\over 2} \int_0^R dr\, r^2 \,\rho\left(\xi_h^2+\xi_z^2\right)=1. 
\label{eq:adia-norm}
\end{equation}
The conservation of the WKB energy flux implies that the region
between each pair of neighboring radial nodes contributes an equal
amount to the above integral.

To achieve a simple analytic result, we take advantage of the
following: $|\xi_h| \gg |\xi_z|$ and $\xi_h\sim \xi_h\big\vert_{\ph}$
for $z\ll z_\omega$, and the envelope of $\rho v_{gz}\xi_h^2\approx$
constant for $z\gg z_\omega$ (cf. eq. [\ref{eq:adia-Fluxp}]).  This
enables us to write
\be
{\omega^2\over 2} R^2 
\xi^2_h\big\vert_{\ph}\left(\rho v_{gz}\right)_\omega\int
{dz\over v_{gz}} \approx 1, \label{eq:adia-normp}
\ee
where the subscript $\omega$ stands for quantities evaluated at
$z_\omega$. Using
\be
\int_0^R {dz\over v_{gz}}\approx {\pi n\over \omega}, 
\label{eq:adia-scaling-eqn} 
\ee
we arrive at     
\be \left(\rho v_{gz}\right)_\omega=\left({\omega^2\rho\over k_h
N}\right)_\omega\sim \left({\omega^2 pz\over
k_h c_s^3}\right)_\omega \sim {k_h^2
F\tau_\omega\over\omega},
\label{eq:adia-scaleing-coeff}
\ee
where $F$ is the stellar flux, and
\be
\tau_\omega\approx\left({pz\over
F}\right)_\omega
\label{eq:adia-scaling-tone} 
\ee
is the thermal time scale at $z_\omega$.

Putting these relations together, we obtain
\be k_h^2\xi_h^2\big\vert_{\ph}\sim {1\over n\tau_\omega L},
\label{eq:adia-scaling-normfin} \ee
where $L=4\pi R^2F$ is the stellar luminosity. Equation (\ref{eq:adia-xihe}) 
then implies 
\be 
\left({{\delta p}\over{p}}\right)^2_{\ph}\sim
\left({{\delta \rho}\over{\rho}}\right)^2_{\ph}\sim {1\over
n\tau_\omega L}. \label{eq:adia-scaling-normdrho} 
\ee
This is an appropriate normalization formula for modes having
$ z_\omega > z_b $.   

\begin{figure} 
\centerline{\psfig{figure=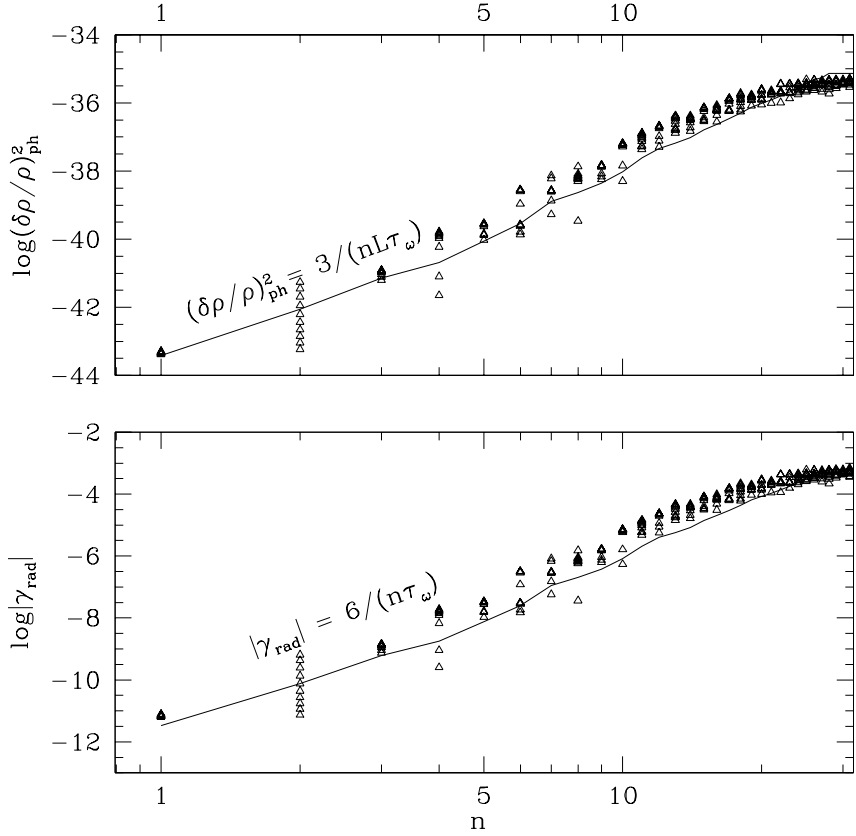,width=0.75\hsize}}
\caption[]{Normalized surface amplitudes as well as driving/damping rates 
for g-modes in white dwarfs.  We compare numerical values obtained
using Bradley's model with $T_{\rm eff} = 12,000 \K$ ($\tau_b = 300
\s$) with analytic estimates. Both properties depend much more
strongly on $n$ than on $\ell$; hence the choice of $n$ as the
abscissa. The upper panel plots as triangles numerical values of
$(\delta\rho/\rho)_{\rm ph}^2$ in $\gm^{-1}{\rm cm}^{-2}
\s^2$ for modes with ${\ell} = 1$ to $10$. 
The eigenfunctions are normalized according to equation
\refnew{eq:adia-norm}. The solid line represents the analytical estimate from 
equation \refnew{eq:adia-scaling-normdrho} for modes with $z_\omega >
z_b$, where $z_\omega$ is chosen to be $3\pi\omega^2/(2gk_h^2)$. The
lower panel displays numerical radiative damping rates as triangles,
together with the analytic estimate from equation
\refnew{eq:adia-Wdotr} as a solid line.
\label{fig:adia-scaling-drhogamma} }
\end{figure}

\end{appendix}



\end{document}